\documentclass[conference,a4paper]{APSIPA2021}
\usepackage{multirow}
\usepackage{graphicx}
\usepackage{amsmath}
\usepackage{amssymb}
\usepackage{amsxtra}
\usepackage{threeparttable}

\usepackage{url} 

\usepackage{bm} 
\renewcommand{\theenumi}{(\alph{enumi}} 
\usepackage{flushend} 
\usepackage{relsize} 
\newcommand{\lsum}{\mathsmaller{\sum}}

\begin{document}

\title{Prior Distribution Design for\\ Music Bleeding-Sound Reduction Based on Nonnegative Matrix Factorization}

\author{%
\authorblockN{%
Yusaku Mizobuchi\authorrefmark{1}, 
Daichi Kitamura\authorrefmark{1}, 
Tomohiko Nakamura\authorrefmark{2}, 
Hiroshi Saruwatari\authorrefmark{2},\\
Yu Takahashi\authorrefmark{3}, 
and Kazunobu Kondo\authorrefmark{3}
}
\authorblockA{%
\authorrefmark{1}
National Institute of Technology, Kagawa College, Kagawa, Japan}
\authorblockA{%
\authorrefmark{2}
The University of Tokyo, Tokyo, Japan}
\authorblockA{%
\authorrefmark{3}
Yamaha Corporation, Shizuoka, Japan}
}

\maketitle
\thispagestyle{empty}

\begin{abstract}
  When we place microphones close to a sound source near other sources in audio recording, the obtained audio signal includes undesired sound from the other sources, which is often called cross-talk or bleeding sound.
  For many audio applications including onstage sound reinforcement and sound editing after a live performance, it is important to reduce the bleeding sound in each recorded signal.
  However, since microphones are spatially apart from each other in this situation, typical phase-aware blind source separation (BSS) methods cannot be used.
  We propose a phase-insensitive method for blind bleeding-sound reduction.
  This method is based on time-channel nonnegative matrix factorization, which is a BSS method using only amplitude spectrograms.
  With the proposed method, we introduce the gamma-distribution-based prior for leakage levels of bleeding sounds. Its optimization can be interpreted as maximum a posteriori estimation.
  The experimental results of music bleeding-sound reduction indicate that the proposed method is more effective for bleeding-sound reduction of music signals compared with other BSS methods.
\end{abstract}

\section{Introduction}
\label{sect:intro}
\begin{figure}[tb]
    \begin{center}
    \vspace{5pt}
        \includegraphics[width=0.8\columnwidth]{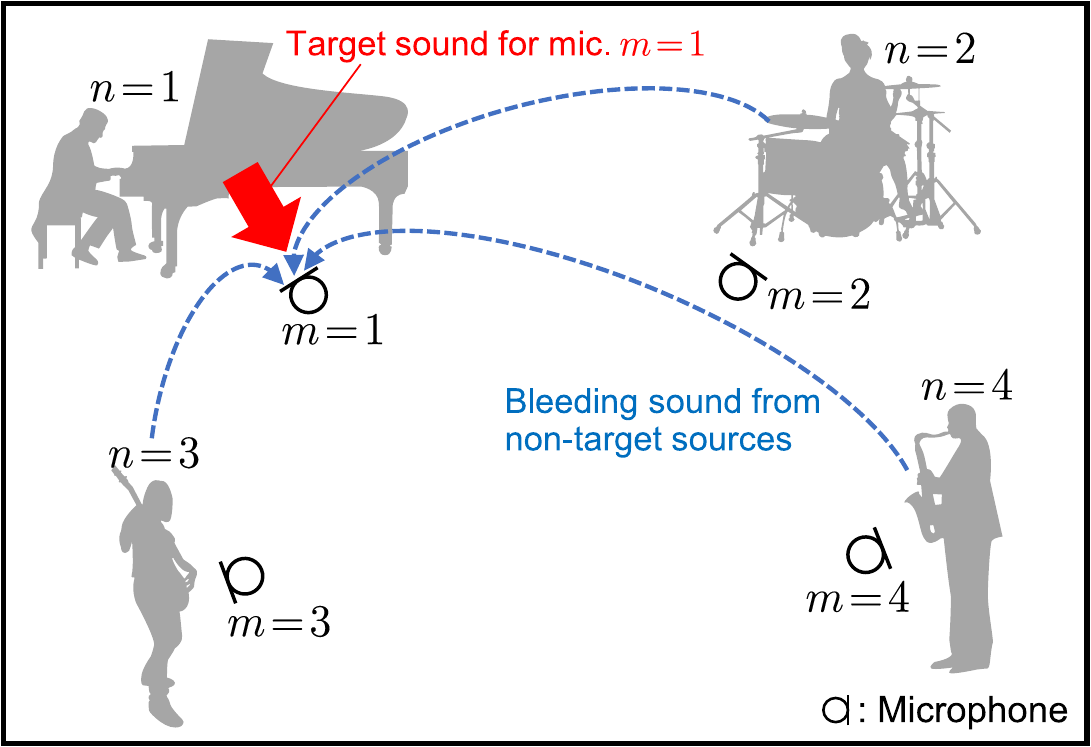}
        \vspace{-10pt}
    \end{center}
    \caption{Spatial arrangement of sources and close microphones, where $M=N=4$. Target sound is contaminated with bleeding sound from other non-target sources.}
    \label{fig:spatialArrangement}
    \vspace{-5pt}
\end{figure}

When we record a live musical performance, many microphones are usually arranged among the players. 
Some are located very close to each of the audio sources, such as musical instruments, vocals, and amplifiers.
These close microphones are placed to pick up the sound from the source other than that which is intended.
However, undesirable audio leakage from the non-target audio sources is also captured, which is often called ``cross-talk'' or ``bleeding sound,'' as shown in Fig.~\ref{fig:spatialArrangement}.

In onstage mixing, sound engineers control the balance of sound levels of individual sources, and the processed sounds are provided to the audience through loudspeakers and performers through monitor speakers. 
Bleeding sound makes such sound reinforcement difficult, degrading musical performance quality.
It is also necessary to avoid sound bleeding for high-quality audio editing (remixing) of the recorded signals after a live performance.
For these reasons, sound engineers carefully place close microphones so that the as much bleeding sound is reduced as possible.
Putting acoustic barriers between the sound sources and reducing sound reflection in the recording room are also effective.
However, completely avoiding bleeding sound is almost impossible.
In other words, sound bleeding essentially occurs in a live-recording situation.

Bleeding-sound reduction is similar to the well-investigated problem called multichannel audio source separation (MASS)~\cite{Brandstein2001_micArrays,VanTrees2002_optArrayProc,Yu2014_bss,Sawada2019_BSS}, but some conditions are different from those in MASS, which are listed as follows.
\begin{enumerate}
    \item The signal-to-noise ratio (SNR) of the observed signal is relatively high because of a close miking setup, where the ``signal'' is a target source for the close microphone and the ``noise'' is the leakage from the other sources.
    \item The observed multichannel signals are already ``labeled,'' namely, the target source for each microphone is known because each microphone is located close to each sound source.
    \item The microphones are spatially apart from each other (e.g., more than 2~m), resulting in serious spatial aliasing.
    \item The requirement of separation quality is relatively high so as not to degrade the artistic value of the music signal.
\end{enumerate}
Conditions (a) and (b) are advantages of bleeding-sound reduction, which make resolving bleeding sound easier than MASS.
However, conditions (c) and (d) are difficult.
In particular, condition (c) is critical because typical high-quality MASS, including beamformers~\cite{Brandstein2001_micArrays,VanTrees2002_optArrayProc} and independence-based blind source separation (BSS)~\cite{Smaragdis1998_fdica,Saruwatari2006_doaPermSolver,Hiroe2006_iva,Kim2007_iva,Ono2011_auxiva,Kitamura2016_ilrma,Kitamura2018_ilrma}, uses phase differences between microphones, which are unreliable in bleeding-sound reduction due to spatial aliasing.
To tackle this problem, phase-insensitive (amplitude- or power-based) MASS~\cite{Togami2010_timeChNmf,Chiba2014_timeChNmf,Murase2014_timeChNmf,Taniguchi2017_dmnmf} can be applied. 
Togami et al.~\cite{Togami2010_timeChNmf} applied nonnegative matrix factorization (NMF)~\cite{Lee1999_nmfNature,Lee2000_nmfEucKl} to the time-channel domain in each frequency (hereafter, time-channel NMF: TCNMF), where both the nonnegative mixing matrix and amplitude activation of each source are estimated in each frequency bin.
TCNMF performs well even under condition (c) or an asynchronous recording condition~\cite{Chiba2014_timeChNmf,Murase2014_timeChNmf}, although its effectiveness regarding music bleeding-sound reduction has not been investigated.
A BSS-based method that ignores the phase information was proposed~\cite{Taniguchi2017_dmnmf}, which is called linear demixed domain multichannel NMF (DMNMF). 
Similar to TCNMF, this method also estimates the frequency-wise nonnegative mixing matrix.
Das et al.~\cite{Das2021_bleedSepMle} introduced supervised information to accurately reduce the bleeding sound, where the frequency-wise nonnegative mixing matrix (i.e., leakage levels of the non-target sources for each close microphone) is measured before the musical performance or calculated using the solo-played time segments of each source.
However, to reduce the onsite recording cost for sound reinforcement, such supervision should not be used.
Also, a mismatch between the obtained mixing matrix and actual condition may markedly degrade reduction performance.

We aimed to reduce bleeding sound in a fully blind manner, namely, the spatial locations of sources and microphones are unknown.
We also did not use supervision of sources, such as solo-played music datasets, to avoid the mismatch between training and test data; thus, supervised deep-neural-network-based approaches~\cite{Nugraha2016_mnmfDnn,Makishima2019_idlma,Kameoka2019_mvae,Makishima2021_idlma,Nakamura2021_mrdla} are out of the scope of this paper.
We propose a phase-insensitive method for blind bleeding-sound reduction, which is a modification of TCNMF: we introduce an a priori generative model for diagonal and off-diagonal elements of the frequency-wise mixing matrix to model relative leakage levels of bleeding sounds.
This method is based on NMF with maximum a posteriori (MAP) estimation, which was originally proposed by Cemgil~\cite{Cemgil_BayesNmf}, and we demonstrate that the proposed method is suitable for reducing music bleeding sound.

\section{Conventional Methods}

\subsection{Mixture Model}

Let $M$ and $N$ be the numbers of microphones (channels) and sources, respectively. 
The source, observed, and estimated signals are respectively denoted as
\begin{align}
    \tilde{\bm{s}}(t) &= [\tilde{s}_1(t), \tilde{s}_2(t), \cdots, \tilde{s}_n(t), \cdots, \tilde{s}_N(t)]^\mathrm{T} \in \mathbb{R}^N, \\
    \tilde{\bm{x}}(t) &= [\tilde{x}_1(t), \tilde{x}_2(t), \cdots, \tilde{x}_m(t), \cdots, \tilde{x}_M(t)]^\mathrm{T} \in \mathbb{R}^M, \\
    \tilde{\bm{y}}(t) &= [\tilde{y}_1(t), \tilde{y}_2(t), \cdots, \tilde{y}_n(t), \cdots, \tilde{y}_N(t)]^\mathrm{T} \in \mathbb{R}^N,
\end{align}
where $t=1,2,\cdots, T$, $n=1,2,\cdots, N$, and $m=1,2,\cdots, M$ are the indices of discrete time, source, and microphone, respectively. 
Under the recording condition described in Sect.~\ref{sect:intro}, the mixing system becomes determined ($M=N$) or overdetermined ($M>N$).
In this study, we focused only on the determined case, which is the most difficult situation in bleeding-sound reduction.

\begin{figure}[tb]
    \begin{center}
    \vspace{5pt}
        \includegraphics[width=0.99\columnwidth]{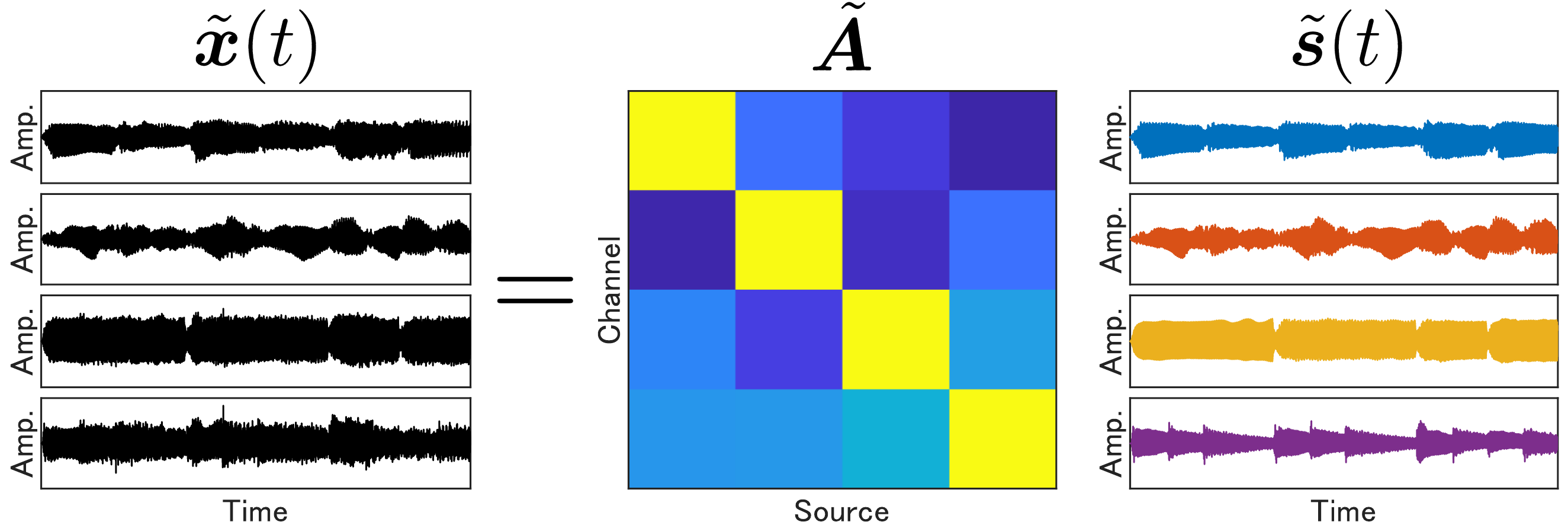}
        \vspace{-20pt}
    \end{center}
    \caption{Instantaneous mixture model for bleeding-sound reduction, where $M=N=4$. Color brightness in mixing matrix $\tilde{\bm{A}}$ shows amplitude level of each element (brighter is larger). Due to close miking setup, diagonal elements in $\tilde{\bm{A}}$ have larger amplitudes compared with off-diagonal elements.}
    \label{fig:obsSrcSig}
    \vspace{-10pt}
\end{figure}

In an instantaneous mixture, the observed and estimated signals can respectively be modeled as 
\begin{align}
    \tilde{\bm{x}}(t) &= \tilde{\bm{A}}\tilde{\bm{s}}(t), \label{eq:anechoicMixture} \\
    \tilde{\bm{y}}(t) &= \tilde{\bm{W}}\tilde{\bm{x}}(t),
\end{align}
where $\tilde{\bm{A}}\in\mathbb{R}^{M\times N}$ and $\tilde{\bm{W}}\in\mathbb{R}^{N\times M}$ are the time-invariant mixing and demixing matrices, respectively.
The mixture model \eqref{eq:anechoicMixture} is illustrated in Fig.~\ref{fig:obsSrcSig}. 
Since the observed signal $\tilde{\bm{x}}(t)$ is ``labeled,'' as explained in condition (b) in Sect.~\ref{sect:intro}, we define that $\tilde{x}_m(t)$ is the close-microphone signal for the $m$th source $\tilde{s}_m(t)$ ($n=m$), as shown in Figs.~\ref{fig:spatialArrangement} and \ref{fig:obsSrcSig}.
Thus, $\tilde{x}_m(t)$ mainly contains the sound from the target source $\tilde{s}_m(t)$, although the bleeding sound from the non-target sources $\tilde{s}_{m'}(t)$ is also included, where $m'\neq m$.
For this reason, the absolute values of diagonal elements in $\tilde{\bm{A}}$ should be large enough, and those of off-diagonal elements become small, which results in high-SNR condition (a) in Sect.~\ref{sect:intro}.
Blind bleeding-sound reduction is formulated as an estimation problem of the mixing matrix $\tilde{\bm{A}}$ or demixing matrix that satisfies $\tilde{\bm{W}}=\tilde{\bm{A}}^{-1}$ from only $\tilde{\bm{x}}(t)$.

In actual recording, the mixing system \eqref{eq:anechoicMixture} becomes a convolutive mixture due to time difference of arrival and room reverberation.
To simply model the convolutive mixture, we assume that the impulse responses (reverberation time) between microphones and sources are shorter than the window length used in the short-time Fourier transform (STFT).
This assumption enables us to respectively model the reverberant observed and estimated signals as\footnote{Note that roman font signal denotes complex values and italic font signal denotes real (or nonnegative) values.} 
\begin{align}
    \mathbf{x}_{ij} &= \mathbf{A}_i\mathbf{s}_{ij} \label{eq:reverbMixture}, \\
    \mathbf{y}_{ij} &= \mathbf{W}_i\mathbf{x}_{ij}, \label{eq:reverbDemixing}
\end{align}
where 
\begin{align}
    \mathbf{s}_{ij} &= [\mathrm{s}_{ij1}, \mathrm{s}_{ij2}, \cdots, \mathrm{s}_{ijn}, \cdots, \mathrm{s}_{ijN}]^\mathrm{T}\in\mathbb{C}^N, \\
    \mathbf{x}_{ij} &= [\mathrm{x}_{ij1}, \mathrm{x}_{ij2}, \cdots, \mathrm{x}_{ijm}, \cdots, \mathrm{x}_{ijM}]^\mathrm{T}\in\mathbb{C}^M, \\
    \mathbf{y}_{ij} &= [\mathrm{y}_{ij1}, \mathrm{y}_{ij2}, \cdots, \mathrm{y}_{ijn}, \cdots, \mathrm{y}_{ijN}]^\mathrm{T}\in\mathbb{C}^N.
\end{align}
Here, $i=1, 2, \cdots, I$ and $j=1, 2, \cdots, J$ are the indices of the frequency bin and time frame, respectively, and $\mathbf{A}_i\in\mathbb{C}^{M\times N}$ is the complex-valued frequency-wise mixing matrix.
Also, $\mathrm{s}_{ijn}$, $\mathrm{x}_{ijm}$, and $\mathrm{y}_{ijn}$ are the complex-valued time-frequency-wise elements of the source, observed, and estimated spectrograms $\mathbf{S}_{n}\in\mathbb{C}^{I\times J}$, $\mathbf{X}_{m}\in\mathbb{C}^{I\times J}$, and $\mathbf{Y}_{n}\in\mathbb{C}^{I\times J}$, respectively.
In \eqref{eq:reverbMixture}, the convolutive mixture is converted to the frequency-wise instantaneous mixture via STFT.

Typical beamformers~\cite{Brandstein2001_micArrays,VanTrees2002_optArrayProc} and BSS methods~\cite{Smaragdis1998_fdica,Saruwatari2006_doaPermSolver,Hiroe2006_iva,Kim2007_iva,Ono2011_auxiva,Kitamura2016_ilrma,Kitamura2018_ilrma} are used to estimate the complex-valued demixing matrix $\mathbf{W}_i$ on the basis of a principle of microphone arrays, e.g., time difference of arrival, and these methods rely on the phase differences between microphones. 
When microphones are spatially apart from each other, these methods cannot precisely estimate $\mathbf{W}_i$ because of spatial aliasing.
This problem is salient in bleeding-sound reduction.

\subsection{DMNMF}

To cope with spatial aliasing, the power-based BSS method DMNMF was proposed~\cite{Taniguchi2017_dmnmf}.
DMNMF can be interpreted as a phase-insensitive version of independent low-rank matrix analysis (ILRMA)~\cite{Sawada2019_BSS,Kitamura2016_ilrma,Kitamura2018_ilrma}, and the observed signal is modeled as 
\begin{align}
    \bm{x}_{ij}^{.2} &\approx \bm{A}_i\bm{s}_{ij}^{.2}\ \ \forall i, j, \\
    \bm{A}_i &= \mathrm{abs}(\mathbf{A}_i)\in\mathbb{R}_{\geq 0}^{M\times N}, \\
    \bm{x}_{ij} &= \mathrm{abs}(\mathbf{x}_{ij})\in\mathbb{R}_{\geq 0}^M, \\
    \bm{s}_{ij} &= \mathrm{abs}(\mathbf{s}_{ij})\in\mathbb{R}_{\geq 0}^N,
\end{align}
where the dotted exponent $\cdot^{.q}$ and absolute operation $\mathrm{abs}(\cdot)$ for vectors or matrices return the element-wise $q$th power and absolute, respectively; thus, $\bm{x}_{ij}^{.2}$ and $\bm{s}_{ij}^{.2}$ are the power spectrogram components of $\{\mathbf{X}_m\}_{m=1}^M$ and $\{\mathbf{S}_n\}_{n=1}^N$, respectively.
DMNMF approximates \eqref{eq:reverbMixture} by the nonnegative frequency-wise mixing matrix $\bm{A}_i$ in the power-spectrogram domain to ignore the phase information. 
In addition, the power spectrogram of each source is modeled by a low-rank matrix using NMF.
After estimating $\bm{A}_i$ and $\bm{s}_{ij}^{.2}$ from $\bm{x}_{ij}^{.2}$, we can recover the estimated signal $\mathbf{y}_{ij}$ by Wiener filtering.

\subsection{TCNMF}

\begin{figure}[tb]
    \begin{center}
    \vspace{5pt}
        \includegraphics[width=0.99\columnwidth]{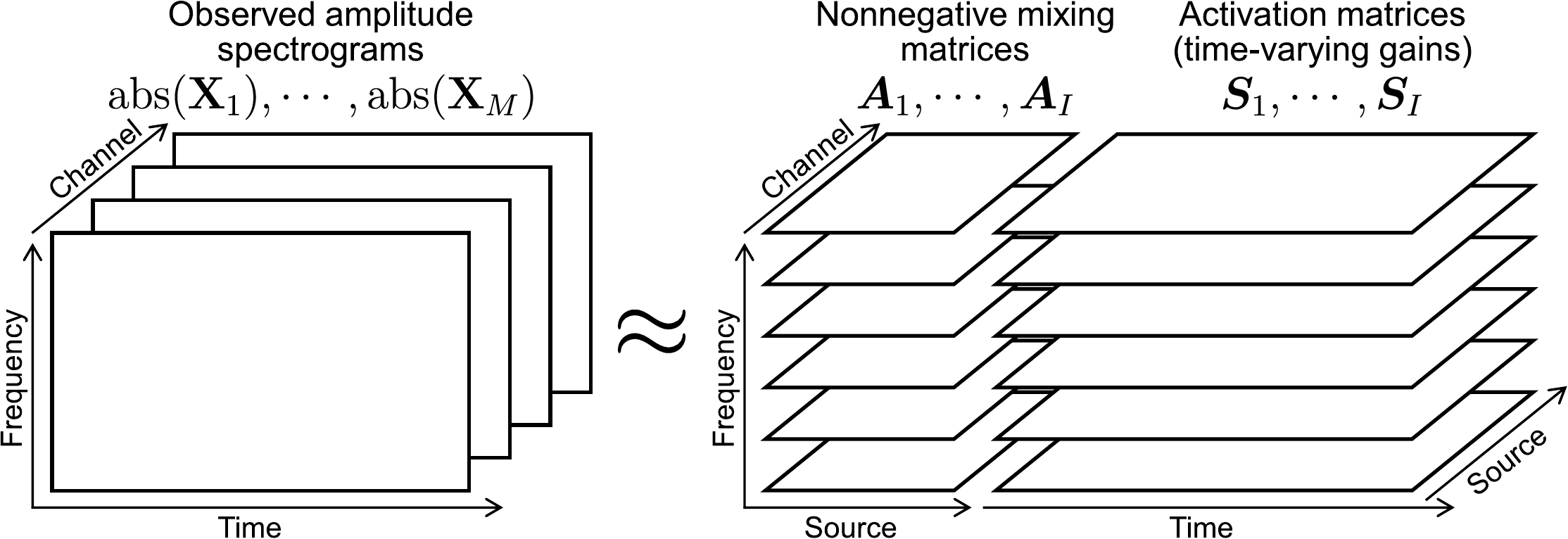}
        \vspace{-25pt}
    \end{center}
    \caption{Decomposition model of TCNMF, where $M=N=4$ and $I = 6$. Note that $\mathrm{abs}(\mathbf{X}_m)$ is channel-wise time-frequency matrix, but $\bm{A}_i$ and $\bm{S}_i$ are frequency-wise source-channel and time-source matrices, respectively.}
    \label{fig:tcnmfModel}
    \vspace{-10pt}
\end{figure}

The amplitude-based BSS method TCNMF was proposed~~\cite{Togami2010_timeChNmf} and applied~\cite{Chiba2014_timeChNmf,Murase2014_timeChNmf} to speech enhancement.
Whereas typical NMF is a low-rank decomposition of time-frequency matrices, TCNMF decomposes frequency-wise time-channel matrices in the amplitude domain as
\begin{align}
    \bm{X}_i &\approx \bm{A}_i\bm{S}_i\ \ \forall i, \\
    \bm{X}_i &= \left[ \bm{x}_{i1}\ \bm{x}_{i2}\ \cdots\ \bm{x}_{iJ} \right]\in\mathbb{R}_{\geq 0}^{I\times J},
\end{align}
which is illustrated in Fig.~\ref{fig:tcnmfModel}, where $\bm{X}_i$ is the frequency-wise time-channel observed signal in the amplitude domain and $\bm{S}_i\in\mathbb{R}_{\geq 0}^{N\times J}$ is a time-source activation matrix: $\bm{S}_i$ involves time-varying gains of each source as the row vectors.
By estimating $\bm{A}_i$ and $\bm{S}_i$ in the same manner as typical NMF, we can reconstruct the estimated sources using Wiener filtering.

The variables $\bm{A}_i$ and $\bm{S}_i$ can be estimated by solving the following minimization problem~\cite{Lee2000_nmfEucKl}:
\begin{align}
    \min_{\mathcal{A}, \mathcal{S}} \sum_i \mathcal{D}_{\mathrm{KL}}\!\left(\bm{X}_i|\bm{A}_i\bm{S}_i\right)~~\mathrm{s.t.}~a_{imn}, s_{inj} \geq 0\ \ \forall i, m, n, j, \label{eq:tcsnmfCost1}
\end{align}
where
\begin{align}
    \nonumber \mathcal{D}_{\mathrm{KL}}\!\left(\bm{X}_i|\bm{A}_i\bm{S}_i\right) &= \sum_{m,j} \left( x_{imj}\log \frac{ x_{imj} }{ \sum_n a_{imn}s_{inj} } \right.\\
    &\left.\phantom{=}\mbox{}- x_{imj} + \sum_n a_{imn}s_{inj} \right) \label{eq:klDiv}
\end{align}
is the generalized Kullback-Leibler (KL) divergence that measures the similarity between $\bm{X}_i$ and $\bm{A}_i\bm{S}_i$, $\mathcal{A}$ and $\mathcal{S}$ are the sets $\{\bm{A}_i\}_{i=1}^I$ and $\{\bm{S}_i\}_{i=1}^I$, respectively, and $x_{imj}$, $a_{imn}$, and $s_{inj}$ are the elements of $\bm{X}_i$, $\bm{A}_i$, and $\bm{S}_i$, respectively.
However, since $\bm{A}_i$ is an $M\times N$ square matrix in the determined case, the minimization problem \eqref{eq:tcsnmfCost1} has a trivial solution, namely, $\bm{A}_i=\bm{I}$ for all $i$, where $\bm{I}$ is an identity matrix.
To avoid this trivial solution, an $L_{0.5}$-norm-based sparse regularizer was introduced for each time frame~\cite{Togami2010_timeChNmf} as follows:
\begin{align}
    \nonumber \min_{\mathcal{A}, \mathcal{S}}\sum_i &\ \mathcal{D}_{\mathrm{KL}}\!\left(\bm{X}_i|\bm{A}_i\bm{S}_i\right) + \mu \sum_{i,j} \| \bm{s}_{ij} \|_{0.5} \\
    \mathrm{s.t.}~&a_{imn}, s_{inj} \geq 0\ \ \forall i, m, n, j, \label{eq:tcsnmfCost2}
\end{align}
where $\mu$ is a weight coefficient for regularization. 
Note that $\bm{s}_{ij}$ is a time-frame-wise vector in $\bm{S}_i$, namely, $\bm{S}_i=[ \bm{s}_{i1}\ \bm{s}_{i2}\ \cdots\ \bm{s}_{iJ} ]$.

\section{Proposed Method}

\subsection{Motivation}

In bleeding-sound reduction, phase information cannot be used because of the close miking setup and serious spatial aliasing.
As a phase-insensitive method, DMNMF is a reasonable approach. 
However, full-blind parameter optimization of DMNMF is difficult and unstable. 
In fact, a priori information of steering vectors (column vectors of $\mathbf{A}_i$) or a phase-aware BSS method is used for pre-estimation~\cite{Taniguchi2017_dmnmf} to stabilize and improve BSS performance. 
TCNMF can estimate the source signals without phase information, even in asynchronous recording~\cite{Chiba2014_timeChNmf}. 
However, its performance for music BSS or bleeding-sound reduction has not been investigated. 
In particular, the sparse regularizer $\sum_{i,j}\|\bm{s}_{ij}\|_{0.5}$ in \eqref{eq:tcsnmfCost2} may degrade the sound quality of estimated signals in music mixture.
This is because the regularizer is based on a W-disjoint-orthogonality assumption in the time-frequency domain~\cite{Yilmaz2004_wDisjoint}, which is suitable only for speech mixtures.
Since music mixtures frequently include both spectral and temporal overlaps of sources, the sparse regularizer for $\bm{S}_i$ may be inappropriate.

To avoid the trivial solution of $\bm{A}_i$ in TCNMF, we use our proposed method to regularize both the diagonal and off-diagonal elements of the nonnegative mixing matrix $\bm{A}_i$ instead of regularizing $\bm{S}_i$.
The proposed method can be interpreted as a MAP estimation, where the bleeding-sound levels are assumed to be generated by the gamma distribution prior.

\subsection{Generative Model of KL-Divergence-Based NMF}

Cemgil~\cite{Cemgil_BayesNmf} revealed the generative model of NMF with KL divergence (KLNMF): the minimization problem in KLNMF is equivalent to the maximum likelihood (ML) estimation with the Poisson generative model.
For \eqref{eq:tcsnmfCost1}, the following generative model is assumed:
\begin{align}
    z_{imnj} &\sim \mathcal{P}(z_{imnj}; a_{imn}s_{inj}), \label{eq:poisson}\\
    \mathcal{P}(z; \lambda) &=\frac{ 1 }{ \Gamma(z+1) }e^{-\lambda}\lambda^{z},
\end{align}
where $z_{imnj}\in\mathbb{N}$ is a random variable that satisfies $x_{imj} = e + \sum_n z_{imnj}$, $\mathcal{P}(z; \lambda)$ is the Poisson distribution with the random variable $z\in\mathbb{N}$ and parameter $\lambda > 0$, $\Gamma(z+1)=z!$ is the gamma function, and $e$ is a random variable that obeys uniform distribution in the range $[0, 1)$. 
Also, $z_{imnj}$ is assumed to be mutually independent w.r.t. $i$, $m$, $n$, and $j$.
The Poisson random variables have the superposition property, namely, when $z_n\sim\mathcal{P}(z_n;\lambda_n)$ and $x=\sum_n z_n$, the marginal probability is given by $p(x) = \mathcal{P}(x; \sum_n \lambda_n)$.
Therefore, the marginal log-likelihood of $\bm{X}_i$ is given by
\begin{align}
    \nonumber \log p(&\bm{X}_i; \bm{A}_i, \bm{S}_i) \\
    \nonumber &= \log \prod_{m, j} \sum_{z_{imnj}} p(x_{inm}; z_{imnj})p(z_{imnj}; a_{imn}s_{inj})\\
    \nonumber &=\log \prod_{m, j} \mathcal{P}(x_{imj}; \lsum_n a_{imn}s_{inj}) \\
    \nonumber &= \sum_{m, j}\left[ x_{imj}\log \sum_n a_{imn}s_{inj} \right.\\
    &\left.\phantom{=}\mbox{}- \sum_n a_{imn}s_{inj} -\log \Gamma(x_{imj}+1) \right]. \label{eq:logLikelihood}
\end{align}
The maximization of \eqref{eq:logLikelihood} w.r.t. $a_{imn}$ and $s_{inj}$ for all $i$ (ML estimation) is equivalent to the minimization of \eqref{eq:klDiv}.

\subsection{A Priori Generative Model for Bleeding-Sound Levels}

\begin{figure}[tb]
    \begin{center}
    \vspace{5pt}
        \includegraphics[width=0.95\columnwidth]{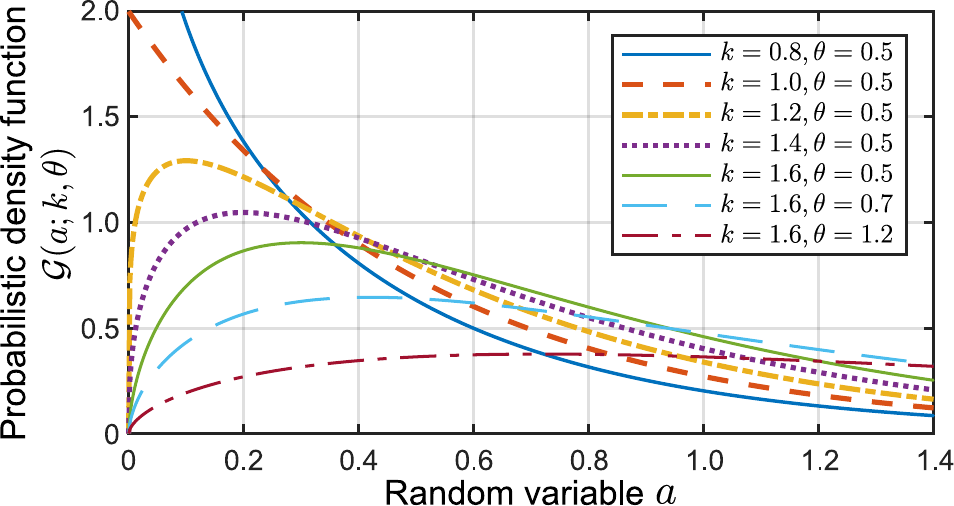}
        \vspace{-15pt}
    \end{center}
    \caption{Probabilistic density function of gamma distribution.}
    \label{fig:gammaPlot}
    \vspace{-14pt}
\end{figure}

With the proposed method, to avoid the trivial solution of $\bm{A}_i$, we introduce the following a priori generative model into the diagonal and off-diagonal elements of $\bm{A}_i$:
\begin{align}
    a_{imn} &\sim 
    \begin{cases}
        \delta(a_{imn}-1)\ \ \ \ \ (m = n) \\
        \mathcal{G}(a_{imn}; k, \theta)\ \ \ \ (m \neq n)
    \end{cases}\!\!\!\!\!, \label{eq:prior} \\
    \mathcal{G}(a; k, \theta) &= \frac{ 1 }{ \Gamma(k)\theta^k }a^{k-1}e^{-a/\theta}, \label{eq:gammaDist}
\end{align}
where $\delta(a)$ is the Dirac's delta distribution and $\mathcal{G}(a; k, \theta)$ is the gamma distribution with the random variable $a\geq 0$ and shape and scale parameters $k>0$ and $\theta>0$. 
Note that the gamma distribution is a conjugate prior of the Poisson generative model \eqref{eq:poisson}.
In addition, $a_{imn}$ is assumed to be mutually independent w.r.t. $i$, $m$, and $n$; thus, the prior distribution of $\bm{A}_i$ becomes 
\begin{align}
    \nonumber p(\bm{A}_i, k, \theta) &= \prod_{m, n=m}p(a_{imn}) \prod_{m, n\neq m}p(a_{imn}; k,\theta) \\
    &= \prod_{m, n=m}\delta(a_{imn}-1) \prod_{m, n\neq m}\mathcal{G}(a_{imn}; k, \theta). \label{eq:priorA}
\end{align}
The prior \eqref{eq:priorA} enables us to control the probability of off-diagonal elements of $\bm{A}_i$ (relative leakage levels of bleeding sound) by $k$ and $\theta$, while restricting all the diagonal elements to be unity.
As shown in Fig.~\ref{fig:gammaPlot}, we can avoid $a_{imn}=0$ for all $m\neq n$, which is the trivial solution of $\bm{A}_i$, by setting the shape parameter to $k>1$. 
Hereafter, we consider $k>1$ only.

For the activation matrix $\bm{S}_i$, we do not assume explicit structure, but only the nonnegativity prior is used as follows:
\begin{align}
    \nonumber s_{inj} &\sim \lim_{\beta\rightarrow\infty} \frac{1}{\beta} \mathcal{I}[0\leq s_{inj}\leq \beta] \\
    &\propto \mathcal{I}[0\leq s_{inj}],
\end{align}
where $\beta$ is the normalized coefficient and  $\mathcal{I}[\cdot]$ denotes a binary distribution that has value one when its argument is true and zero otherwise.
Similar to $\bm{A}_i$, $s_{inj}$ is mutually independent w.r.t. $i$, $n$, and $j$, and the prior distribution of $\bm{S}_i$ becomes 
\begin{align}
    \nonumber p(\bm{S}_i) &= \prod_{n,j}p(s_{inj}) \\
    &\propto \prod_{n,j}\mathcal{I}[0\leq s_{inj}]. \label{eq:priorS}
\end{align}

\subsection{Cost Function for MAP Estimation}

On the basis of the above-mentioned prior distributions, we estimate variables $\bm{A}_i$ and $\bm{S}_i$ in the MAP sense.
The posterior distribution can be obtained as 
\begin{align}
    \prod_i p(\bm{A}_i, \bm{S}_i; \bm{X}_i) \propto \prod_i\underbrace{p(\bm{X}_i; \bm{A}_i, \bm{S}_i)}_{\mathrm{Likelihood}}\underbrace{p(\bm{A}_i, k, \theta)p(\bm{S}_i)}_{\mathrm{Priors}}. \label{eq:posterior}
\end{align}
By taking a negative logarithm of \eqref{eq:posterior}, we can decompose the right side of \eqref{eq:posterior} as 
\begin{align}
    \mathcal{J} = -\sum_i \left[\log p(\bm{X}_i; \bm{A}_i, \bm{S}_i) +\log p(\bm{A}_i, k, \theta) + \log p(\bm{S}_i)\right].
\end{align}
From \eqref{eq:logLikelihood}, \eqref{eq:priorA}, and \eqref{eq:priorS}, the cost function $\mathcal{J}$ is obtained as
\begin{align}
    \nonumber \mathcal{J} &= \sum_{i, m, j}\left[ - x_{imj}\log \sum_n a_{imn}s_{inj} \right.\\
    \nonumber &\left.\phantom{=}\mbox{}+ \sum_n a_{imn}s_{inj} +\log \Gamma(x_{imj}+1) \right] \\
    \nonumber &\left.\phantom{=}\mbox{}+ \sum_{i,m,n=m} \mathbb{I}[a_{imn}=1]\right. \\
    \nonumber &\left.\phantom{=}\mbox{}+ \sum_{i,m,n\neq m} \left[ -(k-1)\log a_{imn} + \frac{1}{\theta}a_{imn} \right]\right. \\
    &\left.\phantom{=}\mbox{}+ \sum_{i,n,j} \mathbb{I}[0\leq s_{imn}]\right.\!, \label{eq:costMap}
\end{align}
where $\mathbb{I}[\cdot] = -\log \mathcal{I}[\cdot]$ denotes an indicator function that has value zero when its argument is true and $\infty$ otherwise.
The MAP estimation of $\bm{A}_i$ and $\bm{S}_i$ is a minimization problem of \eqref{eq:costMap}, and this minimization w.r.t. $\bm{A}_i$ and $\bm{S}_i$ is equivalent to the following problem:
\begin{align}
    \nonumber \min_{\mathcal{A}, \mathcal{S}} \sum_i \mathcal{D}_{\mathrm{KL}}\!\left(\bm{X}_i|\bm{A}_i\bm{S}_i\right) + \sum_{i,m,n\neq m}\mathcal{R}(&a_{imn}; k, \theta) \\
    \mathrm{s.t.}~a_{imn}, s_{inj} \geq 0~\forall i, m, n, j~\mathrm{and}~a_{imn}&=1~\forall m=n,\label{eq:costDiv}
\end{align}
where
\begin{align}
    \mathcal{R}(a_{imn}; k, \theta) = \left[ -(k-1)\log a_{imn} + \frac{1}{\theta}a_{imn} \right]
\end{align}
is the regularizer that corresponds to the gamma distribution prior \eqref{eq:gammaDist} for the off-diagonal elements of $\bm{A}_i$.

\subsection{Derivation of Optimization Algorithm}

The minimization problem \eqref{eq:costDiv} can be solved using a majorization-minimization (MM) algorithm~\cite{Lee2000_nmfEucKl,Sun2017_mmAlgorithmOverview}, which is often used in the context of NMF optimization.
The majorization function of the fidelity term $\mathcal{D}_{\mathrm{KL}}\!\left(\bm{X}_i|\bm{A}_i\bm{S}_i\right)$ can be designed using Jensen's inequality as follows:
\begin{align}
    \nonumber &\mathcal{D}_{\mathrm{KL}}\!\left(\bm{X}_i|\bm{A}_i\bm{S}_i\right) \\ \nonumber &\overset{\mathrm{c}}{=} \sum_{i,m,j} \left( -x_{imj} \log \sum_{n}a_{imn}s_{inj} + \sum_n a_{imn}s_{inj} \right) \\
    \nonumber &= \sum_{i,m,j} \left( -x_{imj} \log \sum_{n} \xi_{imnj}\frac{a_{imn}s_{inj}}{\xi_{imnj}} + \sum_n a_{imn}s_{inj} \right) \\
    \nonumber &\leq \sum_{i,m,j} \left( -x_{imj} \sum_{n} \xi_{imnj} \log \frac{a_{imn}s_{inj}}{\xi_{imnj}} + \sum_n a_{imn}s_{inj} \right) \\
    &\equiv \mathcal{D}^+(\bm{A}_i,\bm{S}_i,\Xi), \label{eq:majorizer}
\end{align}
where $\overset{\mathrm{c}}{=}$ denotes equality up to a constant, $\xi_{imnj}>0$ is an auxiliary variable that satisfies $\sum_n \xi_{imnj} = 1$, and $\Xi$ is a set of $\xi_{imnj}$ for all $i$, $m$, $j$, and $n$. 
The equality in \eqref{eq:majorizer} holds if and only if 
\begin{align}
    \xi_{imnj} = \frac{ a_{imn}s_{inj} }{ \sum_{n'} a_{imn'}s_{in'j}}~\forall i, m, j, n. \label{eq:equalityCond}
\end{align}
From \eqref{eq:majorizer}, the MM problem is obtained as 
\begin{align}
    \nonumber \min_{\mathcal{A}, \mathcal{S}, \Xi} \sum_i \mathcal{D}^+(\bm{A}_i,\bm{S}_i,\Xi) + \sum_{i,m,n\neq m}\mathcal{R}(&a_{imn}; k, \theta) \\
    \nonumber \mathrm{s.t.}~a_{imn}, s_{inj} \geq 0~\forall i, m, n, j,\ \ \xi_{imnj} &> 0~\forall i,m,n,j, \\
    \mathrm{and}~a_{imn}=1~\forall m=n.~~~&\label{eq:costDivMaj}
\end{align}

By setting the derivative of the majorization function \eqref{eq:costDivMaj} w.r.t. $a_{imn}$ and $s_{inj}$ to zero and substituting \eqref{eq:equalityCond} for $\xi_{imnj}$, we can derive the update rules. 
Since the regularizer does not affect $s_{inj}$, the update rule of $s_{inj}$ is the same as that of simple KLNMF~\cite{Lee2000_nmfEucKl} and expressed as 
\begin{align}
    s_{inj} \leftarrow s_{inj} \frac{ \sum_{m} \frac{ x_{imj} }{ \sum_{n'}a_{imn'}s_{inj'} }a_{imn} }{ \sum_{m} a_{imn} }. \label{eq:updateSelement}
\end{align}
For the off-diagonal elements $a_{imn}$ ($m\neq n$), we have the following equations from the derivative of the majorization function:
\begin{align}
    \sum_{j} \left( -x_{imj} \frac{\xi_{imnj}}{a_{imn}} + s_{inj} \right) - (k-1)\frac{1}{a_{imn}} + \frac{1}{\theta} = 0.
\end{align}
Therefore, we have 
\begin{align}
    a_{imn} = \frac{ (k-1) + \sum_j x_{imj}\xi_{imnj} }{ \frac{1}{\theta} + \sum_j s_{inj} }.
\end{align}
The update rule of the off-diagonal elements $a_{imn}$ is derived by substituting \eqref{eq:equalityCond} as
\begin{align}
    a_{imn} \leftarrow \frac{ (k-1) + a_{imn}\sum_j \frac{x_{imj}}{\sum_{n'}a_{imn'}s_{in'j}}s_{inj} }{ \frac{1}{\theta} + \sum_j s_{inj} }. \label{eq:updateAelement}
\end{align}
The nonnegativity of $a_{imn}$ and $s_{inj}$ can hold by setting their initial values to nonnegative values.
Since the value of the diagonal elements of $\bm{A}_i$ is restricted, we initialize the diagonal elements $a_{imn}$ ($m=n$) with unity and fix them during the iterative optimization of the other variables.

The efficient matrix-form implementation of \eqref{eq:updateSelement} and \eqref{eq:updateAelement} is as follows:
\begin{align}
    \bm{A}_i &\leftarrow \frac{ (k-1) + \bm{A}_{i} \odot \left(\frac{\bm{X}_i}{ \bm{A}_i\bm{S}_i} \bm{S}_i^\mathrm{T}\right) }{ \frac{1}{\theta} + \bm{1}\bm{S}_i^\mathrm{T} }\ \ \forall i, \label{eq:updateA1} \\
    \mathrm{diag}(\bm{A}_i) &\leftarrow [1, 1, \cdots, 1]^\mathrm{T}\ \ \forall i, \label{eq:updateA2}\\
    \bm{S}_i &\leftarrow \bm{S}_{i} \odot \frac{ \bm{A}_i^\mathrm{T}\frac{\bm{X}_i}{ \bm{A}_i\bm{S}_i} }{ \bm{A}_i^\mathrm{T}\bm{1} }\ \ \forall i, \label{eq:updateS}
\end{align}
where $\odot$ and the quotient symbol for matrices denote element-wise multiplication and division, respectively, $\bm{1}$ is an $M\times J$ matrix containing only ones, and $\mathrm{diag}(\cdot)$ returns a vector that consists of the diagonal elements of the input square matrix.
Note that \eqref{eq:updateA1} will change the value of the diagonal elements of $\bm{A}_i$, but they are immediately replaced with unity by \eqref{eq:updateA2}.
It is guaranteed that the iterative calculation of \eqref{eq:updateA1}--\eqref{eq:updateS} monotonically decreases the cost function \eqref{eq:costDiv}.

\subsection{Balancing Between Fidelity Term and Regularizer}

With the proposed method, the diagonal elements of $\bm{A}_i$ are restricted to be unity so that the off-diagonal elements correspond to the relative leakage levels of bleeding sound.
The KL divergence \eqref{eq:klDiv} also has a scale-dependent property, namely,
\begin{align}
    \mathcal{D}_{\mathrm{KL}}\!\left(\alpha \bm{X}_i| \alpha \bm{A}_i\bm{S}_i\right) = \alpha \mathcal{D}_{\mathrm{KL}}\!\left(\bm{X}_i|\bm{A}_i\bm{S}_i\right),
\end{align}
where $\alpha\geq 0$ is an arbitrary coefficient.
These facts mean that an observed gain of $\bm{X}_i$, i.e., the signal amplitude in each microphone, affects the balance of the fidelity term $\sum_i\mathcal{D}_{\mathrm{KL}}\!\left(\bm{X}_i|\bm{A}_i\bm{S}_i\right)$ and regularizer $\sum_{i,m,n\neq m}\mathcal{R}(a_{imn}; k, \theta)$ in \eqref{eq:costDiv}.

To solve this problem, we also parameterize the observed gain.
The following normalization is carried out for the observed signal $\tilde{\bm{x}}(t)$ before we apply the proposed method:
\begin{align}
    \tilde{\bm{x}}(t) &\leftarrow \frac{\alpha}{v}\tilde{\bm{x}}(t)\ \ \forall t, \label{eq:preprocess}\\
    v &= \mathrm{max}\!\left(\{\mathrm{abs}(\tilde{\bm{x}}(t))\}_{t=1}^T\right),
\end{align}
where $\mathrm{max}(\cdot)$ returns the maximum scalar value of the input set.
After the normalization \eqref{eq:preprocess}, a dynamic range of $\{\tilde{\bm{x}}(t)\}_{t=1}^T$ becomes $\pm\alpha$.
Similar to $\mu$ in \eqref{eq:tcsnmfCost2}, we can control the balance between the fidelity term and regularizer by $\alpha$.
If we set $\alpha$ to a small value, the regularizer strongly affects the optimization.

\subsection{Reconstruction of Estimated Signals}

Similar to conventional TCNMF, the complex-valued estimated signal $\mathbf{Y}_n$ can be recovered by applying Wiener filtering to the complex-valued observed signal $\mathrm{x}_{ijm}$ as follows:
\begin{align}
    \mathrm{y}_{ijn} = \frac{ (a_{imm}s_{imj})^2 }{ \sum_{n} (a_{imn}s_{inj})^2 } \mathrm{x}_{ijm}. \label{eq:wienerFilt}
\end{align}
Since $a_{imm}=1$, \eqref{eq:wienerFilt} can be implemented as 
\begin{align}
    \mathrm{y}_{ijn} = \left[ \frac{ \bm{S}_i^{.2} }{ \bm{A}_i^{.2}\bm{S}_i^{.2} } \right]_{m,j} \mathrm{x}_{ijm},
\end{align}
where $[\cdot]_{m,j}$ denotes an $(m, j)$ element of the input matrix.
After Wiener filtering, the estimated signal $\mathbf{Y}_n$ is converted to the time-domain signal $\tilde{y}_n(t)$ via the inverse STFT.
Then, the signal gain is recovered by 
\begin{align}
    \tilde{\bm{y}}(t) \leftarrow \frac{v}{\alpha}\tilde{\bm{y}}(t)\ \ \forall t.
\end{align}

\section{Experiments}
\subsection{Conditions}
\label{sect:exp:cond}

To evaluate the performance of the proposed method (proposed TCNMF), we conducted an experiment of blind bleeding-sound reduction.
The observed music mixture signal was simulated using \textit{songKitamura}~\cite{Kitamura2015_hybridNMF,Kitamura_songKitamura}, which is an artificial music dataset. 
We chose four musical instruments, clarinet (Cl.), oboe (Ob.), piano (Pf.), and trombone (Tb), as dry sources $\mathbf{S}_n$ and prepared a four-channel observed signal $\mathbf{x}_{ij}$ so that $M=N=4$. 
To simulate bleeding sound, we mixed these instrumental sounds $\mathbf{s}_{ij}$ using the frequency-wise nonnegative random mixing matrix $\overline{\bm{A}}_i\in\mathbb{R}_{\geq 0}^{M\times N}$ as follows:
\begin{align}
    \mathbf{x}_{ij}=\overline{\bm{A}}_i\mathbf{s}_{ij}, \label{eq:expMixture}
\end{align}
where the diagonal and off-diagonal elements of $\overline{\bm{A}}_i$ are set to unity and uniformly distributed random values in the range $(0,0.2)$ for all $i$, respectively.
This mixing system is an approximation of \eqref{eq:reverbMixture}. 
In this experiment, ten observed mixtures were prepared using different pseudo-random seeds, i.e., ten different mixing matrices $\overline{\bm{A}}_i$.

For all signals, we performed STFT using a 4096-point-long hamming window with half-overlap shifting, where a sampling frequency of the signals was 44.1~kHz. 
The numbers of frequency bins and time frames were $I=2049$ and $J=109$, respectively. 
The update rules in the optimization algorithm were iterated 200 times, and we confirmed the convergence of the cost function value.

For DMNMF and the conventional and proposed TCNMFs, the initial value of $\bm{A}_i$ was set as follows: the diagonal and off-diagonal elements were set to unity and the uniformly distributed random value in the range $(0, 0.1)$, respectively.
The other parameters were initialized by the uniformly distributed random value in the range $(0, 1)$.

As an evaluation criterion, we used the source-to-distortion ratio (SDR)~\cite{Vincent2006_bssEval}, which indicates total separation quality including both degree of separation (source-to-interference ratio: SIR) and absence of artificial distortion (sources-to-artificial ratio: SAR).
As described in condition (a) in Sect.~\ref{sect:intro}, the SNR and SDR of the observed signals for the bleeding-sound reduction are high. 
In our experiment, the average SDRs over the ten observed mixture signals of Cl., Ob., Pf., and Tb. were 18.8, 15.0, 14.7, and 8.6~dB, respectively. 
We calculated the improvements from these input SDRs for each source to evaluate the performance of each method.

We compared five methods, i.e., independent vector analysis (IVA)~\cite{Ono2011_auxiva}, ILRMA~\cite{Kitamura2018_ilrma}, DMNMF~\cite{Taniguchi2017_dmnmf}, the conventional TCNMF~\cite{Togami2010_timeChNmf}, and the proposed TCNMF.
IVA and ILRMA estimate the complex-valued demixing matrix $\mathbf{W}_i$, thus are phase-aware BSS methods. 
The other methods are the phase-insensitive methods that only use amplitude or power spectrograms.
The initial value of $\mathbf{W}_i$ for IVA and ILRMA was set to an inverse matrix of the initial mixing matrix used in DMNMF and the conventional and proposed TCNMFs. 
We also used the numerically stable update rule of the demixing matrix in both IVA and ILRMA, which is called iterative source steering~\cite{Robin2020_iss}, and the estimated source was recovered using \eqref{eq:reverbDemixing}.
We then applied the projection-back technique~\cite{Murata2001_projectionBack} to the estimated signal to recover the frequency-wise signal scales.
For DMNMF and the conventional and proposed TCNMFs, we used Wiener filtering \eqref{eq:wienerFilt} to obtain the estimated source.
For ILRMA and DMNMF, the number of basis vectors in the NMF source model, $L$, was set to $10$, $30$, and $80$.

\begin{figure}[tb]
    \begin{center}
    \vspace{5pt}
        \includegraphics[width=0.95\columnwidth]{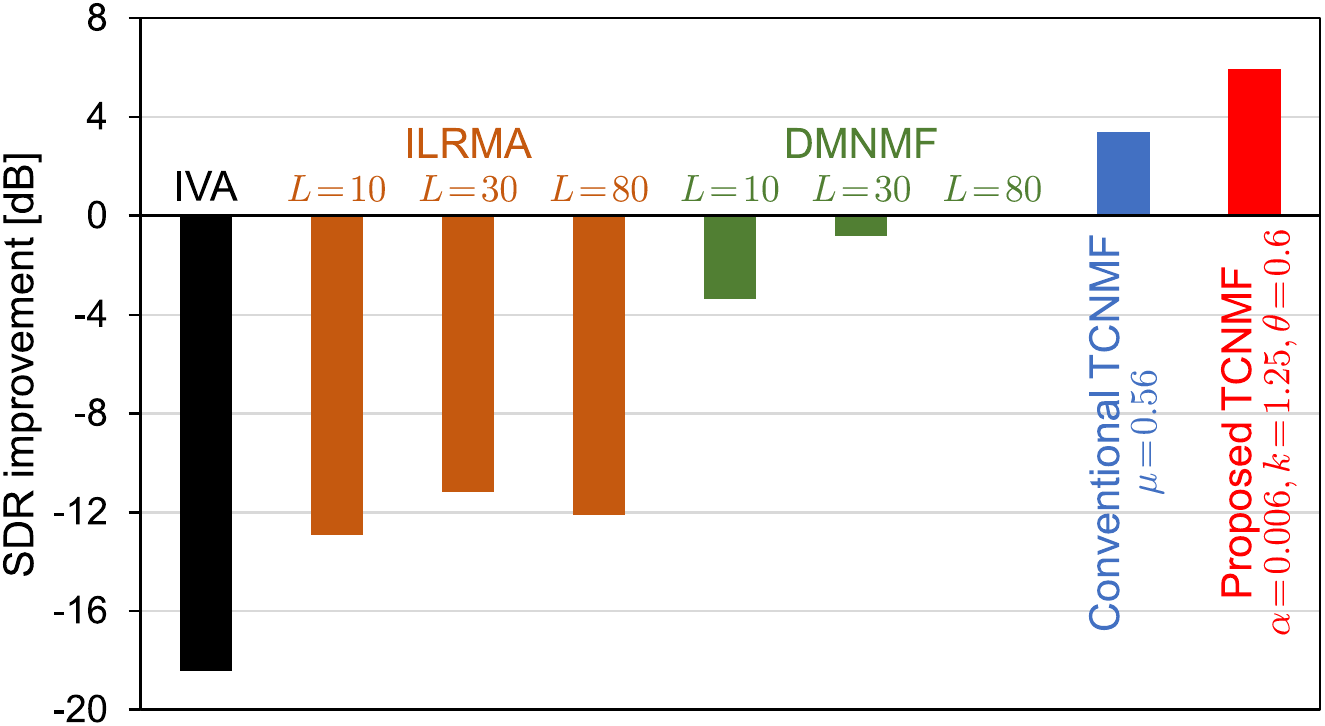}
        \vspace{-12pt}
    \end{center}
    \caption{Comparison of SDR improvements, where each bar is average over 10 different observed mixtures and 4 instrumental sources.}
    \label{fig:resultPlot2}
    \vspace{-10pt}
\end{figure}

\subsection{Results}

Figure~\ref{fig:resultPlot2} shows the average performance comparison among the five methods, where the hyperparameters of the conventional and proposed TCNMF were experimentally determined and set to $\mu=0.56$, $k=1.25$, $\theta=0.6$, and $\alpha=0.006$, which provided the best performance in this experiment. 
We can confirm that the phase-aware BSS methods, IVA and ILRMA, cannot reduce bleeding sound.
This is because the observed mixture signal in this experiment was produced using the nonnegative random mixing matrix $\overline{\bm{A}}_i$ as \eqref{eq:expMixture}, and the phase information is useless for estimating the demixing matrix.
As a result, many artificial distortions are produced in the estimated signals of IVA and ILRMA, degrading their SDR performance.
DMNMF has the potential to reduce bleeding sound, but its performance did not exceed 0~dB.
This result indicates the difficulty of parameter optimization in DMNMF.
For both the conventional and proposed TCNMFs, we can confirm that the average SDR improvements exceed 0~dB. 
In particular, the proposed TCNMF outperformed the conventional TCNMF by more than 2.5~dB. 
This improvement is significant to achieve high-quality post-processing or sound reinforcement of a musical performance.

\section{Conclusion}

We aimed to reduce the bleeding sound in the observed signal obtained with close microphones. 
We proposed a TCNMF method that regularizes the relative leakage levels of bleeding sounds and is based on MAP estimation with the gamma distribution prior.
Experiments using simulated mixture signals showed that the proposed method could achieve the highest bleeding-sound-reduction performance.
Since the proposed method has three hyperparameters, an efficient parameter-tuning method is necessary and is for future work.

\section*{Acknowledgment}

This work was partly supported by JSPS KAKENHI Grant Numbers 19K20306 and 19H01116.


\begin{thebibliography}{99}

  \bibitem{Brandstein2001_micArrays}
  M.~Brandstein and D.~Ward, {\em Microphone Arrays: Signal Processing Techniques and Applications}, Springer-Verlag Berlin Heidelberg, 2001.
  
  \bibitem{VanTrees2002_optArrayProc}
  H.~L.~Van Trees, {\em Optimum Array Processing}, John Wiley and Sons, New York, 2002.

  \bibitem{Yu2014_bss}
  X.~Yu, D.~Hu, J.~Xu, {\em Blind Source Separation: Theory and Applications}, John Wiley and Sons, New York, 2014. 

  \bibitem{Sawada2019_BSS}
  H.~Sawada, N.~Ono, H.~Kameoka, D.~Kitamura, and H.~Saruwatari, 
  ``A review of blind source separation methods: Two converging routes to ILRMA originating from ICA and NMF,'' 
  {\em APSIPA Trans. Signal and Info. Process.}, vol.~8, no.~e12, pp.~1--14, 2019.

  \bibitem{Smaragdis1998_fdica}
  P.~Smaragdis, ``Blind separation of convolved mixtures in the frequency domain,'' {\em Neurocomputing}, vol.~22, pp.~21--34, 1998.
  
  \bibitem{Saruwatari2006_doaPermSolver}
  H.~Saruwatari, T.~Kawamura, T.~Nishikawa, A.~Lee, and K.~Shikano, ``Blind source separation based on a fast-convergence algorithm combining ICA and beamforming,''  {\em IEEE Trans. Audio, Speech, and Lang. Process.}, vol.~14, no.~2, pp.~666--678, 2006.
  
  \bibitem{Hiroe2006_iva}
  A.~Hiroe, ``Solution of permutation problem in frequency domain ICA using multivariate probability density functions,'' {\em Proc. Int. Conf. Independent Compon. Anal. Blind Source Separation}, pp.~601--608, 2006. 

  \bibitem{Kim2007_iva}
  T.~Kim, H.~T.~Attias, S.-Y.~Lee, and T.-W.~Lee, ``Blind source separation exploiting higher-order frequency dependencies,'' {\em IEEE Trans. Audio, Speech, and Lang. Process.}, vol.~15, no.~1, pp.~70--79, 2007.
  
  \bibitem{Ono2011_auxiva}
  N.~Ono, ``Stable and fast update rules for independent vector analysis based on auxiliary function technique,'' {\em Proc. IEEE Workshop Appl. Signal Process. Audio Acoust.}, pp.~189--192, 2011.
  
  \bibitem{Kitamura2016_ilrma}
  D.~Kitamura, N.~Ono, H.~Sawada, H.~Kameoka, and H.~Saruwatari, ``Determined blind source separation unifying independent vector analysis and nonnegative matrix factorization,''  {\em IEEE/ACM Trans. Audio, Speech, and Lang. Process.}, vol.~24, no.~9, pp.~1626--1641, 2016.

  \bibitem{Kitamura2018_ilrma}
  D.~Kitamura, N.~Ono, H.~Sawada, H.~Kameoka, and H.~Saruwatari, ``Determined blind source separation with independent low-rank matrix analysis,'' in  {\em Audio Source Separation}, S.~Makino, Ed., pp.~125--155. Springer, Cham, 2018.

  \bibitem{Togami2010_timeChNmf}
  M.~Togami, Y.~Kawaguch, H.~Kokubo, and Y.~Obuchi, 
  ``Acoustic echo suppressor with multichannel semi-blind non-negative matrix factorization,'' 
  {\em Proc. Asia-Pacific Signal Info. Process. Assoc. Annu. Summit Conf.}, pp.~522--525, 2010.
   
  \bibitem{Chiba2014_timeChNmf}
  H.~Chiba, N.~Ono, S.~Miyabe, Y.~Takahashi, T.~Yamada, and S.~Makino,
  ``Amplitude-based speech enhancement with nonnegative matrix factorization for asynchronous distributed recording,''
  {\em Proc. Int. Workshop Acoustic Signal Enhancement}, pp.~203--207, 2014.
  
  \bibitem{Murase2014_timeChNmf}
  Y.~Murase, H.~Chiba, N.~Ono, S.~Miyabe, Y.~Takahashi, T.~Yamada, and S.~Makino,
  ``On microphone arrangement for multichannel speech enhancement based on nonnegative matrix factorization in time-channel domain,''
  {\em Proc. Asia-Pacific Signal Info. Process. Assoc. Annu. Summit Conf.}, 2014.
  
  \bibitem{Taniguchi2017_dmnmf}
  T.~Taniguchi and T.~Masuda, 
  ``Linear demixed domain multichannel nonnegative matrix factorization for speech enhancement,'' 
  {\em Proc. IEEE Int. Conf. Acoust., Speech Signal Process.}, pp.~476--480, 2017.
   
  \bibitem{Das2021_bleedSepMle}
  O.~Das, J.~O.~Smith, and J.~S.~Abel, 
  ``Microphone cross-talk cancellation in ensemble recordings with maximum likelihood estimation,'' 
  {\em Proc. Audio Eng. Soc. Convention}, 2021.

  \bibitem{Lee1999_nmfNature}
  D.~D.~Lee and H.~S.~Seung,
  ``Learning the parts of objects by non-negative matrix factorization,''
  {\em Nature}, vol.~401, no.~6755, pp.~788--791, 1999.

  \bibitem{Lee2000_nmfEucKl}
  D.~D.~Lee and H.~S.~Seung, 
  ``Algorithms for non-negative matrix factorization'' 
  {\em Proc. Neural Info. Process. Syst.}, pp. 556--562, 2000. 
  
  \bibitem{Nugraha2016_mnmfDnn}
  A.~A.~Nugraha, A.~Liutkus, and E.~Vincent, 
  ``Multichannel audio source separation with deep neural networks,'' 
  {\em IEEE/ACM Trans. Audio, Speech, and Lang. Process.}, vol.~24, no.~9, pp.~1652--1664, 2016.
  
  \bibitem{Makishima2019_idlma}
  N.~Makishima, S.~Mogami, N.~Takamune, D.~Kitamura, H.~Sumino, S.~Takamichi, H.~Saruwatari, and N.~Ono, 
  ``Independent deeply learned matrix analysis for determined audio source separation,'' 
  {\em IEEE/ACM Trans. Audio, Speech, and Lang. Process.}, vol.~27, no.~10, pp.~1601--1615, 2019.
  
  \bibitem{Kameoka2019_mvae}
  H.~Kameoka, L.~Li, S.~Inoue, and S.~Makino, 
  ``Supervised determined source separation with multichannel variational autoencoder,'' 
  {\em Neural Comput.}, vol.~31, no.~9, pp.~1891--1914, 2019.
  
  \bibitem{Makishima2021_idlma}
  N.~Makishima, Y.~Mitsui, N.~Takamune, D.~Kitamura, H.~Saruwatari, Y.~Takahashi, and K.~Kondo, 
  ``Independent deeply learned matrix analysis with automatic selection of stable microphone-wise update and fast sourcewise update of demixing matrix,'' 
  {\em Signal Process.}, vol.~178, 107753, 2021.
  
  \bibitem{Nakamura2021_mrdla}
  T.~Nakamura, S.~Kozuka, and H.~Saruwatari, 
  ``Time-domain audio source separation with neural networks based on multiresolution analysis,''
  {\em IEEE/ACM Trans. Audio, Speech, and Lang. Process.}, vol.~29, pp. 1687--1701, 2021.

  \bibitem{Cemgil_BayesNmf}
  A.~T.~Cemgil,
  ``Bayesian inference for nonnegative matrix factorisation models,''
  {\em Computational Intelligence and Neuroscience}, vol.~2009, no.~785152, 2009.

  \bibitem{Yilmaz2004_wDisjoint}
  O.~Y{\i}lmaz and S.~Rickard, ``Blind separation of speech mixtures via time-frequency masking,'' {\em IEEE Trans. Signal Process.}, vol.~52, no.~7, pp. 1830--1847, 2004.

  \bibitem{Sun2017_mmAlgorithmOverview}
  Y.~Sun, P.~Babu, and D.~P.~Palomar, 
  ``Majorization-minimization algorithms in signal processing, communications, and machine learning,''
  {\em IEEE Trans. Signal Process.}, vol.~65, no.~3, 2017.

  \bibitem{Kitamura2015_hybridNMF}
  D.~Kitamura, H.~Saruwatari, H.~Kameoka, Y.~Takahashi, K.~Kondo, and S.~Nakamura, 
  ``Multichannel signal separation combining directional clustering and nonnegative matrix factorization with spectrogram restoration,'' 
  {\em IEEE/ACM Trans. Audio, Speech, and Lang. Process.}, vol.~23, no.~4, pp.~654--669, 2015.
  
  \bibitem{Kitamura_songKitamura}
  D.~Kitamura, 
  ``Open dataset: songKitamura,'' 
  \url{http://d-kitamura.net/dataset_en.html}. Accessed 10 July 2021.
  
  \bibitem{Vincent2006_bssEval}
  E.~Vincent, R.~Gribonval, and C.~F\'evotte, 
  ``Performance measurement in blind audio source separation,'' 
  {\em IEEE Trans. Audio, Speech, and Lang. Process.}, vol.~14, no.~4, pp.~1462--1469, 2006.

  \bibitem{Robin2020_iss}
  S.~Robin and N.~Ono, 
  ``Fast and stable blind source separation with rank-1 updates,'' 
  {\em Proc. IEEE Int. Conf. Acoust., Speech Signal Process.}, pp.236--240, 2020.
  
  \bibitem{Murata2001_projectionBack}
  N.~Murata, S.~Ikeda, and A.~Ziehe,
  ``An approach to blind source separation based on temporal structure of speech signals,''
  {\em Neurocomputing}, vol.~41, no.~1--4, pp.~1--24, 2001.

\end{thebibliography}
\end{document}